\documentstyle[aps,prl,twocolumn]{revtex}

\begin{document}
\narrowtext
\title{ Comment on ``Consistent Sets Yield Contrary Inferences in Quantum Theory''}

\author{Robert B. Griffiths}
\address{Department of Physics,
Carnegie Mellon University,
Pittsburgh, PA 15213}

\author{James B. Hartle}
\address{Institute for Theoretical Physics,
University of California,
Santa Barbara, CA 93108}

\maketitle

	In \cite{kt97}, Kent correctly points out that in consistent histories
quantum theory it is possible, given initial and final states, to construct two
different consistent families of histories (what Kent calls consistent sets),
in each of which there is a proposition that can be inferred with probability
one, and such that the projectors representing these two propositions are
mutually orthogonal.  In response we stress that, according to the rules of
consistent history reasoning {\it two such propositions are not contrary in the
usual logical sense} \cite{cn93}: namely, that one can infer that if one is
true then the other is false, and both could be false. No single consistent
family contains {\it both} propositions, together with the initial and final
states, and hence the propositions cannot be logically compared.

	Let us say projectors $P$ and $Q$ are ``perpendicular'', to employ a
term with no logical connotation, if $PQ=0$ and $P\neq I-Q$.  If two
``perpendicular'' projectors occur {\it in the same consistent family}, they
represent mutually exclusive events such that if the probability of one is 1
(true), that of the other is 0 (false), and thus the events corresponding to
these projectors are ``contrary'' in the logical sense.  However, the
requirement stated in italics is absolutely essential, for it is a basic rule
of reasoning in the consistent history formalism that all inferences must be
carried out in a single consistent family \cite{om94,gr96}.

	As an example \cite{av91}, suppose a particle can be in one of three
non-overlapping boxes, for which the corresponding orthogonal states are
$|A\rangle, |B\rangle, |C\rangle,$ and assume the dynamics is trivial: if the
particle starts in one box, it stays there.  Let
\begin{equation} 
	\begin{array}{c}
	|\Phi\rangle = (|A\rangle + |B\rangle + |C\rangle)/\sqrt 3,\\
	|\Psi\rangle = (|A\rangle + |B\rangle - |C\rangle)/\sqrt 3,
	\end{array}
\label{e1}
\end{equation}
be initial and final states at times $t_0$ and $t_2$.  There is a consistent
family of histories ${\cal A}$ including these initial and final states, and,
at a time $t_1$ between $t_0$ and $t_2$, the projectors $A= |A\rangle{\langle}
A|$ and $\tilde A = I-A$, which correspond to the particle being or not being
in box $A$.  Application of the consistent history formalism yields
$\Pr(A|\Phi,\Psi)=1$; that is, using family $\cal A$, it is certain that the
particle at time $t_1$ is in box $A$.  There is also a second consistent family
${\cal B}$ in which $B= |B\rangle{\langle} B|$ and $\tilde B = I-B$ are
possible intermediate states at time $t_1$, and using this family one finds
$\Pr(B|\Phi,\Psi)=1$.

	The projectors $A$ and $B$ are ``perpendicular'' in the sense defined
above.  However, there is no consistent family which includes $\Phi$ at $t_0$
and $\Psi$ at $t_2$ together with {\it both} $A$ and $B$ at $t_1$. Thus
inferences that concern both $A$ and $B$, such as that the events are contrary,
cannot be drawn from the initial and final data.  (For additional remarks, see
App.~A of \cite{gr97}.)
	Kent employs the term ``contrary'' for the relationship between
projectors that we have been calling ``perpendicular'', whether or not they
occur in the same consistent family.  But importing a term from classical logic
into quantum theory must be done with some care in order to avoid confusion,
as illustrated by the example just discussed.

	By contrast, if a projector $P$ occurs in a consistent family, $I-P$
necessarily occurs in the same family, according to the rules in \cite{gr96}.
Hence, one {\it can} say that $P$ and $I-P$ are ``contradictory''
in the sense of classical logic \cite{cn93}.  There are no initial and
final states for which projectors $P$ and $I-P$ can have probability one in
different consistent families, because the probability assigned to any event on
the basis of given data is independent of the consistent
family in which it occurs.  Kent \cite{kt97} expresses concern that the
consistent history formalism thereby treats contradictory and perpendicular
projectors in different ways. However, we see no necessary reason for symmetry,
because, while a pair $P$ and $I-P$ can always be associated with the logical
notion of ``contradictory'', perpendicular projectors need not be associated
with the logical concept of ``contrary'', as in the examples Kent is concerned
with.  Alternative, more restrictive, formulations of consistent histories
quantum mechanics, as in \cite{kt97b}, should be
judged as theories on their own merits and ultimately by comparison with
experiment.

	Consistent histories quantum theory is logically consistent, consistent
with experiment, consistent with the usual quantum predictions for
measurements, and applicable to the most general physical systems.  It may not
be the only theory with these properties, but in our opinion, it is the most
promising among present possibilities.

	The research of RG has been supported in part by the
National Science Foundation through grant PHY 96-02084.
The work of JH was supported in part by NSF grants PHY 95-07065 and PHY
94-07194.

\end{document}